\documentclass[pra,twocolumn,showpacs,preprintnumbers,amsmath,amssymb,floatfix]{revtex4}
\usepackage[dvips]{graphicx}
\usepackage{dcolumn}
\usepackage{bm}

\allowdisplaybreaks[2]

\begin{document}

\title{Magnetic response enhancement via electrically induced magnetic moments}

\author{Bastian \surname{Jungnitsch}}
\email{bastian.jungnitsch@mpi-hd.mpg.de}
\author{J\"org \surname{Evers}}
\email{joerg.evers@mpi-hd.mpg.de}
\affiliation{Max-Planck-Institut f\"ur Kernphysik, Saupfercheckweg 1, D-69117 Heidelberg}

\date{\today}

\begin{abstract}
The realization of negative refraction in atomic gases requires a strong magnetic response of the atoms. Current proposals for such systems achieve an enhancement of the magnetic response by a suitable laser field configuration, but still rely on high gas densities. Thus further progress is desirable, and this requires an
understanding of the precise mechanism for the enhancement.
Therefore, here we study the magnetic and electric response to a probe field interacting 
with three-level atoms in ladder configuration. In our first model, the three transitions are driven by a control field and the electric and magnetic component of the probe field, giving rise to a closed interaction loop. In a reference model, the coherent driving is replaced by an incoherent pump field. A time-dependent analysis of the closed-loop system enables us to identify the different contributions to the medium response. A comparison with the reference system then allows one to identify the physical mechanism that leads to the enhancement. It is found that the enhancement occurs at so-called multiphoton resonance by a scattering of the coupling field and the electric probe field mode into the magnetic probe field mode. Based on these results, conditions for the enhancement are discussed.
\end{abstract}

\maketitle

\section{Introduction}
As Veselago pointed out in 1968, materials having both negative permittivity and permeability can acquire a negative index of refraction~\cite{veselago,review1}. Such negative index materials are also called ``left-handed'', since the electric ($\vec{E}$) and magnetic ($\vec{H}$) components of an electromagnetic wave travelling trough a negatively refracting medium and its wave vector form a left-handed coordinate system.
These materials offer promising applications~\cite{D.R.Smith08062004,Sh2007,review1}, such as the possibility to overcome the diffraction limit with a negatively refracting, perfect lens, as proposed by Pendry~\cite{pendry:lens}. Therefore it is not surprising that in the recent past, left-handed materials and negative refraction have been studied intensely. These efforts have been fueled by a multitude of successful experimental demonstrations of negative refraction and related effects, mostly relying on metamaterials~\cite{R.A.Shelby04062001,PhysRevLett.90.137401,CostasM.Soukoulis01052007,zhang:137404,soukoulis, eleftheriades}. Metamaterials are artificial structures with feature size below the incident radiation wavelength that allow to control the electromagnetic response to a great extent and at the same time appear as a bulk medium to the incident radiation.

A different ansatz is the quantum optical realization of negative refraction in dense atomic gases~\cite{oktel, mandel, fleischhauer,orth:paper}. Negative refraction, however, typically requires both electric and magnetic response at the probe field frequency, which is hampered by the fact that usually the coupling of the magnetic component of a probe laser field to a magnetic dipole transition in atoms is strongly suppressed. A simple order-of-magnitude estimate shows that the suppression factor is proportional to two powers of the finestructure constant, $\alpha^2\sim 137^{-2}$.
Thus, additional effort is required in order to enhance the magnetic response. In the literature, a number of schemes that allow to achieve a high positive index of refraction with small absorption have been proposed. They are based on a suitable modification of the electric response of the medium (see, e.g.,~\cite{PhysRevA.46.1468}). The enhancement, however, is typically too small such that a direct transfer of these ideas to magnetic transitions is not straightforward. Thus, the schemes suggested so far for negative refraction rely on a different mechanism that is related to an enhancement via chirality~\cite{pendry:chirality,oktel,fleischhauer,orth:paper}. The medium is such that the magnetic response is influenced by both the electric and the magnetic component of the probe field and analogously for the electric response. A first interpretation for the case of atomic systems has been given in~\cite{oktel}, which however did not consider the coupling of the magnetic probe field component to the atomic system in deriving the induced magnetic dipole moment, and which focusses on a certain resonance case for the applied fields and the employed level structure. Also, the system considered there only allows for enhancement at a single frequency, while an enhancement at a range of probe field frequencies was reported in subsequent work~\cite{fleischhauer,orth:paper}.
Thus better insight is desirable, not least since it might lead to further enhancement of the magnetic response which would significantly simplify the theoretical and experimental study of negative refraction in atomic gases. 

Motivated by this, here we study in detail the enhancement mechanism that is at the heart of current schemes to achieve negative refraction in atomic gases. For this, we revisit the three-level ladder-type system studied in~\cite{oktel}, where one transition is driven by a coherent control field and the other two transitions couple to the magnetic and electric component of a probe field. In contrast to previous studies, we apply a time-dependent analysis of the medium response that enables us to directly identify the various physical processes contributing to the medium response. These results are compared to a reference system that is obtained by replacing the coherent driving field by an incoherent pumping field. We find that the enhancement of the magnetic response occurs since the used level schemes are so-called closed-loop media. In these systems, the laser fields applied are such that they form a closed interaction loop. We identify a scattering of the coupling field and of the electric probe field into the magnetic probe field component as the mechanism responsible for the enhancement of the response and provide conditions for this process to take place. It is found that the so-called multiphoton resonance condition must be fulfilled for the enhancement to be present. In the studied three-level system, the condition is satisfied only at a single probe field frequency. But in larger level schemes, the laser fields can be applied in such a way that the enhancement works at arbitrary frequencies of the probe field.

The paper is organized as follows: In Sec.~\ref{sec2}, we present our model systems and derive the equations of motion as well as expressions for the medium response coefficients. In Sec.~\ref{analytical_results}, we solve the equations of motion, both in the time-dependent case for the closed-loop configuration, for the time-independent case at multiphoton resonance and in the incoherently pumped system. Using these results, in Sec.~\ref{compare} we compare the different systems and finally identify the enhancement mechanism. Sec.~\ref{discussion} discusses and summarizes the results.


\section{\label{sec2}Theoretical considerations}
\subsection{Model}
\begin{figure}[t]
\centering
\includegraphics[width=0.45\columnwidth]{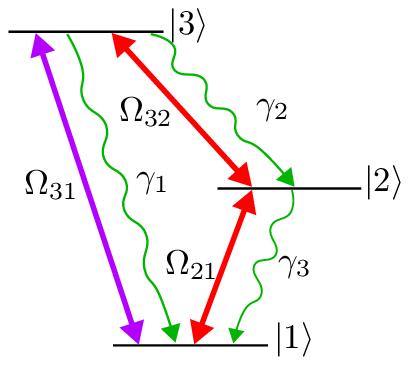}
\includegraphics[width=0.45\columnwidth]{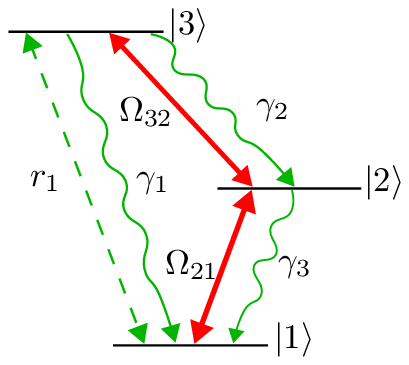}
\caption{\label{fig:closed_loop_scheme}(Color online) (a) The three-level 
system driven by coherent fields in loop configuration. The probe field
components are denoted by red solid double arrows, the coupling field
by purple solid double arrows. Spontaneous emission is indicated by the 
wiggly green arrows. The transition $|1\rangle \leftrightarrow |2\rangle$
couples to the magnetic component, while transition $|2\rangle \leftrightarrow
|3\rangle$ couples to the electric component of the same probe field. 
(b) Reference system obtained by replacing the coherent control field 
by an incoherent, bi-directional pumping, indicated by the green 
dashed double arrow.}
\end{figure}

We start by writing down the applied electromagnetic fields as shown in  Fig.~\ref{fig:closed_loop_scheme}. Since we treat both systems semi-classically, we have
\begin{subequations}
\label{fields}
\begin{eqnarray}
\label{el_field}
\vec{E}(\vec{r},t) = \: && \vec{E}_{p}(\vec{r}) e^{i \phi} e^{-i \omega_p t} + \vec{E}_{c}(\vec{r}) e^{i \psi} e^{-i \omega_{c} t} \nonumber
\\
&& + \: \textrm{c.c.} \:,
\\
\label{mag_field}
\vec{B}(\vec{r},t) = \: && \vec{B}_{p}(\vec{r}) e^{i \phi} e^{-i \omega_p t} + \vec{B}_{c}(\vec{r}) e^{i \psi} e^{-i \omega_{c} t} \nonumber
\\
&& + \: \textrm{c.c.} \:,
\end{eqnarray}
\end{subequations}
where the subindex p [c] refers to the probe [control] field.
Further, $\vec{E}_{p}(\vec{r}) = \mathcal{E}_p\vec{e}_{p} e^{i \vec{k}_{p} \vec{r}}$, where $\mathcal{E}_{p}$ is the electric field amplitude, $\vec{e}_p$ the unit polarization vector of the electric component of the probe field, $\vec{k}_p$ is the probe field's wave vector, $\omega_p$ its frequency and $\phi$ its absolute phase.
Analogously, $\vec{E}_{c}(\vec{r}) = \mathcal{E}_{c} \vec{e}_c e^{i \vec{k}_{c} \vec{r}}$, where $\mathcal{E}_{c}$ is the control field amplitude and $\vec{e}_c$ its unit polarization vector. $\vec{k}_c$ is the wave vector of the control field, $\omega_c$ its frequency and $\psi$ the control field's total phase. The magnetic probe field component is defined analogously as
$\vec{B}_{p}(\vec{r}) = \mathcal{B}_p \vec{b}_{p} e^{i \vec{k}_{p} \vec{r}}$. Note that the magnetic probe field component unit polarization vector is $\vec{b}_p = \vec{\kappa}\times \vec{e}_p$ with unit propagation direction vector $\vec{\kappa} = \vec{k}_p/k_p$. 

In rotating-wave and dipole approximation, we arrive at the Hamiltonian
\begin{subequations}
\label{Hamiltonian}
\begin{align}
H = & H_{0} + H_{I}\,, 
\\
H_0 = & \sum\limits_{j = 1}^{3} \hbar \omega_{j} \vert j \rangle \langle j \vert\,, \\ 
H_I = & - \hbar \: \left(\Omega_{21} e^{-i\omega_{p} t}  \vert 2 \rangle \langle 1 \vert + \Omega_{32} e^{-i\omega_{p} t}  \vert 3 \rangle \langle 2 \vert \right. \nonumber
\\
& \left. +\: \Omega_{31} e^{-i\omega_{c} t}  \vert 3 \rangle \langle 1 \vert + \textrm{H.c.} \right) \: .
\end{align}
\end{subequations}
Note that in the rotating wave approximation, the magnetic control field component can be neglected, because it does not couple near-resonantly to a magnetic transition.
In Eqs.~(\ref{Hamiltonian}), the energy of state $|i\rangle$ is $\hbar \omega_i$ ($i\in\{1,2,3\}$) and the Rabi frequencies are defined as 
\begin{subequations}
\begin{align}
\Omega_{21} &= e^{i \phi} \vec{B}_{p} (\vec{r})\: \vec{\mu}_{21} / \hbar\,,\\ \Omega_{32} &= e^{i \phi} \vec{E}_{p} (\vec{r})\: \vec{d}_{32} / \hbar\,,\\
\Omega_{31} &= e^{i \psi} \vec{E}_{c} (\vec{r})\: \vec{d}_{31}/ \hbar\,.
\end{align}
\end{subequations}
The electric dipole moments are defined as $\vec{d}_{32} = \langle 3 \vert \vec{d} \vert 2 \rangle$ and $\vec{d}_{31} = \langle 3 \vert \vec{d} \vert 1 \rangle$. Analogously, the magnetic dipole moment is $\vec{\mu}_{21} = \langle 2 \vert \vec{\mu} \vert 1 \rangle$, with $\vec{\mu}_{12} = \vec{\mu}_{21}^*$. Here, $\vec{d}$ and $\vec{\mu}$ are the electric and magnetic dipole operator, respectively, and $\vec{d}_{ij} = \vec{d}_{ji}^*$, $\vec{\mu}_{ij} = \vec{\mu}_{ji}^*$, and $\Omega_{ij} = \Omega_{ji}^*$.

\subsection{Equations of motion}
In this section, we derive the equations of motion for a general system that contains both systems of interest shown in Fig. \ref{fig:closed_loop_scheme} as special cases.
To this end, we consider a three-level system that - on transition $|1\rangle \leftrightarrow |2\rangle$ - combines coherent pumping by a control field and bi-directional incoherent pumping. Since this system is a closed-loop system, in general there is no stationary state in the long-time limit and the Hamiltonian necessarily has an intrinsic explicit time dependence~\cite{Buckle,Kosachiov,PhysRevLett.70.3243,hinze,PhysRevA.53.3444,PhysRevA.59.2302,PhysRevA.60.4996,PhysRevLett.84.5308,PhysRevA.66.053409,kajari-schroder:013816,PhysRevLett.93.223601,PhysRevLett.93.190502,shpaisman:043812,evers}.
To simplify this time dependence, we apply the transformation
\begin{equation}
\label{transformation}
V = e^{\frac{i}{\hbar} (H_{0} + X) t}  (H_{I} - X) e^{-\frac{i}{\hbar} (H_{0} + X) t} \: ,
\end{equation}
where $X = \Delta_{1} \vert 1 \rangle \langle 1 \vert + \Delta_{2} \vert 2 \rangle \langle 2 \vert$. Here, we chose the notations $\Delta_{1} = \omega_{3} - \omega_{1} - \omega_{c}$ and $\Delta_{2} = \omega_{3} - \omega_{2} - \omega_{p}$ for the detunings. By applying this transformation, we arrive at the following equations of motions, if we include spontaneous decay in the Born-Markov approximation:
\begin{subequations}
\label{eom}
\begin{align}
\frac{\partial}{\partial t} \tilde{\varrho}_{11}  = &  
- \:r_1 \tilde{\varrho}_{11} + \gamma_3 \tilde{\varrho}_{22} 
+ \left(\gamma_1 + r_1\right) \tilde{\varrho}_{33} 
+ i e^{-i\Delta t } \Omega_{12} \tilde{\varrho}_{21} 
\nonumber \allowdisplaybreaks[2] \\
& - \: i e^{i\index{}\Delta t} \Omega_{21} \tilde{\varrho}_{12} 
+ i \Omega_{13} \tilde{\varrho}_{31} 
- i \Omega_{31} \tilde{\varrho}_{13} \: ,
\allowdisplaybreaks[2] \\
\frac{\partial}{\partial t} \tilde{\varrho}_{12}  = & 
-\:\left[i \left(\Delta - \Delta_3\right) 
+ \frac{1}{2} \left(r_1+\gamma_3\right)\right] \tilde{\varrho}_{12}  
- i  \Omega_{32} \tilde{\varrho}_{13} 
\nonumber \allowdisplaybreaks[2] \\
& - \:i e ^{-i \Delta t} \Omega_{12} \left(\tilde{\varrho}_{11}
  - \tilde{\varrho}_{22}\right) +  i \Omega_{13} \tilde{\varrho}_{32} \: ,
\allowdisplaybreaks[2] \\
\frac{\partial}{\partial t} \tilde{\varrho}_{13}  = &  
- \:\left[i\left(\Delta - \Delta_2 -  \Delta_3\right) 
+\frac{1}{2} \left(2 r_1 + \gamma_1  
+\gamma_2\right)  \right] \tilde{\varrho}_{13} 
\nonumber \allowdisplaybreaks[2] \\
&- \: i \Omega_{13} \left( \tilde{\varrho}_{11} 
-  \tilde{\varrho}_{33}\right) 
- i  \Omega_{23} \tilde{\varrho}_{12} 
+ i e^{- i \Delta t} \Omega_{12} \tilde{\varrho}_{23} \: ,
\allowdisplaybreaks[2] \\
\frac{\partial}{\partial t} \tilde{\varrho}_{22}  = & 
- \:\gamma_3  \tilde{\varrho}_{22} 
+ \gamma_2  \tilde{\varrho}_{33} 
- i e^{-i \Delta t} \Omega_{12}  \tilde{\varrho}_{21} 
\nonumber \allowdisplaybreaks[2] \\
&+\: i e^{i \Delta t} \Omega_{21}  \tilde{\varrho}_{12}
 - i \Omega_{32}  \tilde{\varrho}_{23}
 + i \Omega_{23}  \tilde{\varrho}_{32} \: ,
\allowdisplaybreaks[2] \\
\frac{\partial}{\partial t} \tilde{\varrho}_{23}  = & 
- \:\left[-i \Delta_2 + \frac{1}{2} \left(r_1 
+ \gamma_1  + \gamma_2 + \gamma_3 \right)\right] \tilde{\varrho}_{23} 
\nonumber \allowdisplaybreaks[2] \\
&- i \Omega_{13} \tilde{\varrho}_{21} 
+ \: i e^{i \Delta t} \Omega_{21} \tilde{\varrho}_{13} 
+ i \Omega_{23} \left(\tilde{\varrho}_{33} 
-\tilde{\varrho}_{22}\right) \: ,
\allowdisplaybreaks[2] \\
\tilde{\varrho}_{33}  = & 1 - \tilde{\varrho}_{11} - \tilde{\varrho}_{22} \: .
\end{align}
\end{subequations}
Here, $\tilde{\varrho}_{ij} \: (i,j \: \in \: \lbrace1,2,3\rbrace)$ is 
the density matrix in the interaction picture obtained by transformation of $\varrho_{ij}$ according to Eq. (\ref{transformation}).  One can see that this transformation  simplifies the explicit  time dependence on the right hand side of the equations of motion to factors of $e^{\pm i \Delta t}$ in front of the weak magnetic probe field Rabi frequency $\Omega_{21}$ or $\Omega_{12}$, respectively. 
$\gamma_i$ are spontaneous emission rates on the different transitions.
Also, we introduced the detuning on the magnetic probe field transition $\Delta_{3} = \omega_{2} - \omega_{1} - \omega_{p}$, as well as
the so-called multiphoton detuning 
\begin{equation}
\label{detuning_equality}
\Delta = \Delta_2 + \Delta_3 - \Delta_1 \: ,
\end{equation}
which for the current system evaluates to
\begin{equation}
\label{detuning_explicit}
\Delta = \omega_p - 2 \omega_c \,.
\end{equation}
By setting the incoherent pumping rate $r_1 = 0$ in Eqs.~(\ref{eom}), we arrive at the equations of motion for the system in Fig.~\ref{fig:closed_loop_scheme}(a). Setting 
$\Omega_{31} = \Omega_{13}= \Delta_3 = 0$ yields the equations of motion for the incoherently pumped system shown in Fig.~\ref{fig:closed_loop_scheme}(b).

\subsection{Electric and magnetic responses}
Since our aim is to study the magnetic and electric responses, we require an expression for them in terms of the density matrix elements governed by Eqs.~(\ref{eom}). We will find such a relation in this subsection. For this, it is important to note that electric fields can not only induce electric polarization, but also magnetization~\cite{pendry:chirality,fleischhauer}. Similarly, magnetic fields can induce both magnetization and polarization.

For definitiveness, in the following we specialize to a circularly polarized ($\sigma^{+}$) probe field and probe field propagation in $z$ direction, since one then obtains for the probe field polarization vectors $\vec{b}_{p} = - i \vec{e}_{p}$, i.e., they are parallel. Therefore, the tensorial structure
of the response coefficients in the macroscopic polarization $\vec{P}$ and 
magnetization $\vec{M}$ simplifies considerably,
\begin{subequations}
\label{macro_pol}
\begin{eqnarray}
\vec{P}(\vec{r}, t) = && \frac{1}{c} \int^{\infty}_{-\infty} \xi_{EH}(\tau) \: \vec{H}(\vec{r}, t - \tau) \: d\tau \nonumber
\\
&& + \: \varepsilon_0 \int^{\infty}_{-\infty} \chi_e(\tau) \: \vec{E}(\vec{r}, t - \tau) \: d\tau \:,
\\
\vec{M}(\vec{r}, t) = && \frac{1}{c \mu_0} \int^{\infty}_{-\infty} \xi_{HE}(\tau) \: \vec{E}(\vec{r}, t - \tau) \: d\tau \nonumber
\\
&& + \int^{\infty}_{-\infty} \chi_m(\tau) \: \vec{H}(\vec{r}, t - \tau) \: d\tau \:,
\end{eqnarray}
\end{subequations}
 and the electric and magnetic susceptibility $\chi_e$ and $\chi_m$ and the chirality coefficients $\xi_{HE}$ and $\xi_{EH}$ become scalars. While the susceptibilities determine the electric [magnetic] response to the electric [magnetic] probe field component, the chiralities or cross-terms determine the magnetic response to the electric probe field component and vice versa. Here, $c$ is the vacuum speed of light, and $\mu_0$ and $\varepsilon_0$ are the vacuum permeability and permittivity.
Note that, with the notation of Eqs.~(\ref{macro_pol}), the refractive index in Fourier space is given as~\cite{fleischhauer}
\begin{align}
\label{refractive_index}
n(\omega) =& \sqrt{\tilde{\varepsilon}(\omega) \tilde{\mu}(\omega)  - \frac{1}{4} \left[\tilde{\xi}_{EH}(\omega) + \tilde{\xi}_{HE}(\omega)\right ]^2} \nonumber\\
&+ \frac{i}{2}\left[\tilde{\xi}_{EH}(\omega) - \tilde{\xi}_{HE}(\omega)\right ] \,,
\end{align}
where 
\begin{subequations}
\begin{align}
\tilde{\varepsilon}(\omega) = \tilde{\chi}_e(\omega) + 1 \,,\\
\tilde{\mu}(\omega) = \tilde{\chi}_m(\omega) + 1 \,,
\end{align}
\end{subequations}
are the permittivity $\tilde{\varepsilon}(\omega)$ and the permeability $\tilde{\mu}(\omega)$ in Fourier space. Also, $\tilde{\xi}_{EH}(\omega)$, $\tilde{\xi}_{HE}(\omega)$, $\tilde{\chi}_{m}(\omega)$ and $\tilde{\chi}_{e}(\omega)$ are the Fourier transformed response coefficients of the corresponding time domain quantities in Eqs. (\ref{macro_pol}).  They satisfy $\tilde{\chi}_e(\omega)^* = \tilde{\chi}_e(-\omega)$ and similar for the other susceptibilities and the chiralities. For vanishing chirality coefficients, Eq.~(\ref{refractive_index}) reduces to the well-known relation between the refractive index and the permittivity and permeability for non-chiral media~\cite{jackson}.

Let us now continue to find an expression of the response coefficients in terms of the density matrix elements. Plugging Eqs.~(\ref{fields}) into Eqs.~(\ref{macro_pol}), we arrive at 
\begin{subequations}
\label{macro_pol_Fourier}
\begin{align}
\vec{P} = & \: \frac{\tilde{\xi}_{EH}(\omega_p)}{c \mu_0} \vec{B}_p(\vec{r}) e^{i (\phi  -\omega_p t)} 
\nonumber \\
&+ \: \frac{ \tilde{\xi}_{EH}(\omega_c)}{c \mu_0} \vec{B}_c(\vec{r}) e^{i (\psi -\omega_c t)} 
\nonumber \\
&  + \: \varepsilon_0 \tilde{\chi}_e(\omega_p) \vec{E}_p(\vec{r}) e^{i (\phi -\omega_p t)} 
\nonumber \\
&+ \: \varepsilon_0 \: \tilde{\chi}_e(\omega_c) \vec{E}_c(\vec{r}) e^{i (\psi  -\omega_c t)}+ \: \textrm{c.c.} \:,
\\
\vec{M} = & \: \frac{\tilde{\xi}_{HE}(\omega_p)}{\mu_0 c} \vec{E}_p(\vec{r}) e^{i (\phi -\omega_p t)} 
\nonumber \\
&+ \: \frac{\tilde{\xi}_{HE}(\omega_c)}{\mu_0 c} \vec{E}_c(\vec{r}) e^{i (\psi -\omega_c t)} 
\nonumber \\
& + \: \frac{1}{\mu_0} \tilde{\chi}_m(\omega_p) \vec{B}_p(\vec{r}) e^{i (\phi -\omega_p t)} 
\nonumber \\
&+ \: \frac{1}{\mu_0} \tilde{\chi}_m(\omega_c) \vec{B}_c(\vec{r}) e^{i (\psi -\omega_c t)} + \: \textrm{c.c.} \:.
\end{align}
\end{subequations}
Here, we have used $\vec{B} = \mu_0 \vec{H}$, which holds since $\vec{B}$ is an external field. Note that the  Rabi frequencies in the equations of motion contain local fields, whereas Eqs.~(\ref{macro_pol}) contain external fields~\cite{orth:paper}. However, local field effects are not considered here and therefore we do not distinguish between external and local fields, as is valid for moderate particle densities.

The polarization is also given by $\vec{P} = N \vec{p}$, and the magnetization by $\vec{M} = N \vec{m}$, where $N$ is the particle density and $\vec{p}$ and $\vec{m}$ are the mean polarization and magnetization per atom~\cite{ScZu1997}.
We can express the mean polarization as $\vec{p} = Tr(\varrho \vec{d})$ and the mean magnetization as $\vec{m} = Tr(\varrho \vec{\mu})$. These traces can be written in terms of the transformed density matrix elements $\tilde{\varrho}_{ij} \: (i,j \: \in \: \lbrace1,2,3\rbrace)$ which are given as solutions of the equations of motion (\ref{eom}). 
If the light travels through the medium over a macroscopic distance, then the magnetic and electric response is determined by the part of the medium response that oscillates in phase with the probe field. In order to identify the relevant parts, we apply another unitary transformation into a system oscillating in phase with the probe field. We denote the density matrix in this frame by $\hat{\varrho}_{ij} \: (i,j \: \in \: \lbrace1,2,3\rbrace)$. 

In the incoherently pumped system of Fig.~\ref{fig:closed_loop_scheme}(b), the Hamiltonian is time-independent, and we find
\begin{subequations}
\label{coh_equality}
\begin{eqnarray}
\hat{\varrho}_{21} & = & \tilde{\varrho}_{21} \: ,
\\
\hat{\varrho}_{32} & = & \tilde{\varrho}_{32} \: .
\end{eqnarray}
\end{subequations}
On the other hand, in the closed-loop system of Fig.~\ref{fig:closed_loop_scheme}(a), the coherences $\hat{\varrho}_{32}$ and $\hat{\varrho}_{21}$ are given by
\begin{subequations}
\label{second_trafo}
\begin{eqnarray}
\hat{\varrho}_{21} & = & e^{i \omega_p t} \varrho_{21} 
 =  e^{-i(\omega_c - 2 \omega_p)t} \tilde{\varrho}_{21} \: ,
\\
\hat{\varrho}_{32} & = & e^{i \omega_p t} \varrho_{32} 
 = \tilde{\varrho}_{32} \: .
\end{eqnarray}
\end{subequations}
One can see that at multiphoton resonance $\Delta=0 \Leftrightarrow \omega_c = \omega_p/2$ [see Eq.~(\ref{detuning_explicit})], the two reference frames denoted by $\hat{\rho}_{ij}$ and $\tilde{\rho}_{ij}$ coincide for the two coherences in Eq.~(\ref{second_trafo}),
similar to the case of incoherent pumping. We will investigate the multiphoton resonance case further in Sec.~\ref{multiphoton_resonance}.

We now proceed with the evaluation of the response coefficients. Keeping only terms oscillating in phase with the probe field, we arrive at
\begin{subequations}
\label{macro_response}
\begin{align}
\tilde{\chi}_e & =  \frac{N}{\varepsilon_0 \hbar} \: d_{32}^{2} \: \hat{\varrho}_{32}^{(0,1)}  \, ,
\\
\tilde{\chi}_m & =   \frac{N \mu_0}{\hbar}  \: \mu_{21}^{2} \: \hat{\varrho}_{21}^{(1,0)} \, ,
\\
\tilde{\xi}_{HE} & =  -i\: \frac{N c \mu_0 }{\hbar} \: d_{32} \: \mu_{21} \: e^{i \Phi} \hat{\varrho}_{21}^{(0,1)} \,,
\\
\tilde{\xi}_{EH} & =  i\: \frac{N c\mu_{0}}{\hbar} \:  d_{32} \: \mu_{21} \: e^{-i \Phi} \hat{\varrho}_{32}^{(1,0)} \,,
\end{align}
\end{subequations}
where the $\hat{\varrho}_{ij}^{(1,0)}$ and $\hat{\varrho}_{ij}^{(0,1)}$ are  expansion coefficients in a Taylor series of $\hat{\varrho}_{ij} \: (i,j \: \in \: \lbrace1,2,3\rbrace)$ in terms of $\Omega_{32}$ and $\Omega_{21}$:
\begin{subequations}
\label{coherence_expansion}
\begin{align}
\hat{\varrho}_{32} = &\: \hat{\varrho}_{32}^{(0,0)} + \hat{\varrho}_{32}^{(0,1)} \Omega_{32} + \hat{\varrho}_{32}^{(1,0)} \Omega_{21} \nonumber
\\
& + O(\Omega_{21}^2, \Omega_{32}^2, \Omega_{21}\Omega_{32}) \: ,
\\
\hat{\varrho}_{21} = & \: \hat{\varrho}_{21}^{(0,0)} + \hat{\varrho}_{21}^{(0,1)} \Omega_{32} + \hat{\varrho}_{21}^{(1,0)} \Omega_{21} \nonumber
\\
& + O(\Omega_{21}^2, \Omega_{32}^2, \Omega_{21}\Omega_{32}) \: .
\end{align}
\end{subequations}
In Eqs.~(\ref{coherence_expansion}), we call $\hat{\varrho}_{32}^{(0,1)} \Omega_{32}$ and $\hat{\varrho}_{21}^{(1,0)} \Omega_{21}$ the ``direct terms'', since they correspond to the susceptibilities, while $\hat{\varrho}_{32}^{(1,0)} \Omega_{21}$ and $\hat{\varrho}_{21}^{(0,1)} \Omega_{32}$  are denoted ``cross terms'' as they give rise to the chiralities.
Also, in Eqs.~(\ref{macro_response}), we have introduced the relative phase,
\begin{align}
\Phi = \phi_{32}-\phi_{21}\,,
\end{align}
between the scalar dipole moments which we write as
\begin{align}
\vec{d}_{32} \vec{e}_{p} &= d_{32}\,e^{i\phi_{32}}\,,\\
\vec{\mu}_{21} \vec{b}_{p} &= \mu_{21}\,e^{i\phi_{21}}\,.
\end{align}
Eqs.~(\ref{macro_response}) are the desired relation between the density matrix elements and the coefficients that determine the magnetic response. These can now be used in Eqs.~(\ref{macro_pol_Fourier}) in order to determine the contribution of the various processes to the polarization and magnetization. Keeping only the terms relevant to the probe field response in phase with the probe field frequency, we find
\begin{subequations}
\label{final_pol}
\begin{align}
\label{final_elpol}
\vec{P} = &   \frac{N}{\hbar}\: \vec{e}_p\:e^{i (\vec{k}_p \vec{r}  -\omega_p t + \phi)}
\left ( 
 d_{32}  \mu_{21}\:
\mathcal{B}_p \: e^{-i \Phi}\:\hat{\varrho}_{32}^{(1,0)}    \right.\nonumber
\\ &\left. + \: d_{32}^{2} \: \mathcal{E}_p\: \hat{\varrho}_{32}^{(0,1)}  \right) + \textrm{c.c.} \: ,
\\
\label{final_mag}
\vec{M} = &  
\frac{N}{\hbar} \:\vec{b}_p \: e^{i (\vec{k}_p \vec{r}  -\omega_p t + \phi)}
\left (
 d_{32}  \mu_{21} \: \mathcal{E}_p \: e^{i\Phi} \: \hat{\varrho}_{21}^{(0,1)} \right . \nonumber
\\ &  \left . + \: \mu_{21}^{2} \: \mathcal{B}_p \: \hat{\varrho}_{21}^{(1,0)} \right ) + \textrm{c.c.}\: .
\end{align}
\end{subequations}

\section{\label{analytical_results}Analytical results}

We next solve the time-dependent equations of motion~(\ref{eom}) for both of our systems. First, we consider the closed-loop system for arbitrary multiphoton detuning $\Delta$ and derive an expression for the coherences to first order in the magnetic and electric probe field Rabi frequencies $\Omega_{21}$ and $\Omega_{32}$. Then, we consider the special cases $\Delta \neq 0$ and $\Delta = 0$. Finally, we solve the incoherently pumped system.

\subsection{\label{closed_loop_system}Closed-loop system}

The equations of motion for the closed-loop system are obtained from Eqs.~(\ref{eom}), if we set $r_1 = 0$. We define the vector $\tilde{R}$
containing all density matrix elements,
\begin{equation}
\tilde{R} = (\tilde{\varrho}_{11}, \tilde{\varrho}_{12}, \tilde{\varrho}_{13}, \tilde{\varrho}_{21}, \tilde{\varrho}_{22}, \tilde{\varrho}_{23}, \tilde{\varrho}_{31}, \tilde{\varrho}_{32})^{T} \: .
\end{equation}
The equations of motion~(\ref{eom}) can be rewritten in terms of $\tilde{R}$ as
\begin{equation}
\label{R_motion}
\frac{\partial}{\partial t} \tilde{R} = M \tilde{R} + \Sigma \: .
\end{equation}
Here, we have eliminated $\tilde{\varrho}_{33}$ via the trace condition $Tr(\tilde{\varrho}) = 1$, which is the reason for the appearance of the constant term $\Sigma$ in Eq.~(\ref{R_motion}).
We proceed by splitting both M and $\tilde{R}$ up into terms with different time-dependencies as follows:
\begin{eqnarray}
\label{M_notation}
M = && M_{0} + M_{1} \Omega_{21} e^{i \Delta t} + M_{-1} \Omega_{12} e^{-i \Delta t}  \: ,
\\
\label{S_notation}
\Sigma = && \Sigma_{0} + \Sigma_{1} \Omega_{21} e^{i \Delta t} + \Sigma_{-1} \Omega_{12} e^{-i \Delta t}  \: ,
\end{eqnarray}
where $M_{k}$ and $\Sigma_{k}  \: (k \: \in \: \lbrace-1,0,1\rbrace)$ are time-independent. 

According to Floquet's theorem \cite{floquet}, the solution of $\tilde{R}$ has only contributions oscillating with frequencies that are integer multiples of $\Delta$. Since terms oscillating at higher frequencies are suppressed by powers of the magnetic probe field Rabi frequency $|\Omega_{21}|$, we expand $\tilde{R}$ to first order in this Rabi frequency, and work with the ansatz
\begin{equation}
\label{R_ansatz}
\tilde{R} = \tilde{R}_{0} + \tilde{R}_{1} \Omega_{21} e^{i \Delta t} + \tilde{R}_{-1} \Omega_{12} e^{-i \Delta t} + O(|\Omega_{12}|^2) \: .
\end{equation}
From Eqs.~(\ref{R_motion}-\ref{R_ansatz}), a comparison of coefficients yields
\begin{subequations}
\label{R_solution}
\begin{eqnarray}
\label{R_0_solution}
\tilde{R}_{0} = && - M_{0}^{-1} \Sigma_{0} \: ,
\\
\tilde{R}_{1} = && -(M_{0}- i \Delta)^{-1} (M_{1} \tilde{R}_{0} + \Sigma_{1}) \: ,
\\
\tilde{R}_{-1} = && -(M_{0}+ i \Delta)^{-1} (M_{-1} \tilde{R}_{0} + \Sigma_{-1}) \: .
\end{eqnarray}
\end{subequations}
Since the density matrix element $\hat{\varrho}_{21}$ oscillating in phase with the probe field and  $\tilde{\varrho}_{21}$ are related as
\begin{eqnarray}
\hat{\varrho}_{21} = && e^{ -i (\omega_{c} - 2 \omega_{p}) t} \tilde{\varrho}_{21} \nonumber 
\\
= && e^{ -i (\omega_{c} - 2\omega_{p}) t}\: [\tilde{R}_{0}]_4 +  \:[\tilde{R}_{1}]_4 \Omega_{21} \nonumber
\\
&& + \: e^{ -2 i( \omega_{c}-2  \omega_{p}) t} \:[\tilde{R}_{-1}]_4 \Omega_{12} \: ,
\end{eqnarray}
one can determine $[\tilde{R}_{1}]_4$ as the part of $\hat{\varrho}_{21}$ oscillating in phase with the probe field, where the index 4 denotes the fourth component of $\tilde{R}_{1}$. Likewise, the relevant part of $\hat{\varrho}_{32}$ can be identified with $[\tilde{R}_{0}]_8$.

Note that $\tilde{R}_{0}$, $\tilde{R}_{-1}$ and $\tilde{R}_{1}$ are independent of $\Omega_{21}$, but do depend on the electric probe field Rabi frequency $\Omega_{32}$, which we did not take into account so far. Since we are only interested in the linear magnetic and electric response, we still have to expand the appropriate components of $\tilde{R}$ in $\Omega_{32}$.
We obtain
\begin{subequations}
\label{hat_coh_gen}
\begin{align}
\hat{\varrho}_{21} = & [\tilde{R}_{1}]_4 \: \Omega_{21} 
+ [\tilde{R}_{0}]_4\: e^{-i( \omega_c -2 \omega_p) t} \nonumber
\\
& + O(\Omega_{21}^2, \Omega_{32}^2, \Omega_{21}\Omega_{32}) \: ,
\\
\hat{\varrho}_{32} = & [\tilde{R}_{-1}]_8 \: e^{-i(\omega_c - 2\omega_p)t } \Omega_{12} + [\tilde{R}_{0}]_8  \nonumber
\\
& + O(\Omega_{21}^2, \Omega_{32}^2, \Omega_{21}\Omega_{32}) \: ,
\end{align}
\end{subequations}
where $[\tilde{R}_{0}]_4 \propto \Omega_{23}$ and $[\tilde{R}_{0}]_8 \propto \Omega_{32}$, and higher orders of $\Omega_{32}$ have been neglected. 
With Eqs.~(\ref{R_solution}) and (\ref{hat_coh_gen}), an explicit evaluation yields
\begin{widetext}
\begin{subequations}
\label{Floquet_results}
\begin{align}
\hat{\varrho}_{21} = &  \frac{2}{B} \left\lbrace\ \frac{8 \Delta_1^3 \gamma_3 + 4 i \Delta_1^2 \gamma_3 \left(2 i \Delta_3+ \gamma_{s}\right) + \Delta_3 \left[ 8 \vert \Omega_{31} \vert ^2 \left(\gamma_2 - \gamma_3\right)- 2 \Gamma\right] }{4 \vert \Omega_{31} \vert^2 + \left(2 \Delta_3 - i \gamma_3\right)\left(2 \Delta_1 - 2 \Delta_3 + i \gamma_{s}\right)} \right. \nonumber
\allowdisplaybreaks[2]\\
&\left. \qquad + \: \frac{ 2 \Delta_1 \left[-4 \vert \Omega_{31} \vert ^2 \left(\gamma_2 - 2 \gamma_3\right) + \Gamma\right] - i \left[\left(4 \vert \Omega_{31} \vert ^2 \gamma_2 - \Gamma\right) \gamma_{s}  - 4 \vert \Omega_{31} \vert ^2 \gamma_3^2\right]}{4 \vert \Omega_{31} \vert^2 + \left(2 \Delta_3 - i \gamma_3\right)\left(2 \Delta_1 - 2 \Delta_3 + i \gamma_{s}\right)}\right\rbrace \Omega_{21} \nonumber
\allowdisplaybreaks[2]\\
&-\left \lbrace \:    \frac{4}{B} \frac{4 \vert \Omega_{31} \vert ^2 \left(\gamma_2 - \gamma_3\right) +\left( 2  \Delta_1 + i(\gamma_1 + \gamma_2)\right) \gamma_3 \left(-2  \Delta_2 -i \gamma_{s}\right)}{4 \vert \Omega_{31} \vert ^2  + \left[ 2 i \left(\Delta_1 - \Delta_2 \right) + \gamma_3 \right] \left(- 2 i \Delta_2 + \gamma_{s}\right)} \right \rbrace \:e^{-i (\omega_c - 2 \omega_p) t} \:\Omega_{31}\Omega_{23} \, ,
\allowdisplaybreaks[2]\\
\hat{\varrho}_{32} =& \left \lbrace    \frac{4}{B} \frac{4 \vert \Omega_{31} \vert ^2 \left(- \gamma_2 + \gamma_3 \right) + \left[2 \Delta_1 + i \left(\gamma_1 + \gamma_2\right)\right] \gamma_3 \left(2 \Delta_1 - 2 \Delta_3 - i \gamma_{s}\right)}{4 \vert \Omega_{31} \vert ^2 + \left(2 \Delta_3 + i \gamma_3 \right) \left( 2 \Delta_1 - 2 \Delta_3 - i \gamma_{s}\right)} \right \rbrace\: e^{-i(\omega_c - 2\omega_p)t }\: \Omega_{31} \Omega_{12} \: \nonumber
\allowdisplaybreaks[2]\\
& -\left\lbrace  \vert \Omega_{31} \vert ^2 \frac{8}{B}  \frac{\left[2 \left( - \Delta_1 + \Delta_2 \right) \gamma_2 + i \gamma_3 \left(2 i \Delta_2 + \gamma_1 + \gamma_3\right)\right]}{4 \vert \Omega_{31} \vert ^2 + \left(- 2 i \Delta_1 + 2 i \Delta_2 + \gamma_3\right)\left(2 i \Delta_2 + \gamma_{s}\right)} \right\rbrace \:\Omega_{32}\: ,
\end{align}
\end{subequations}
\end{widetext}
where 
\begin{subequations}
\begin{eqnarray*}
B & = & 4 \Delta_1^2 \gamma_3 + \Gamma + 4 \vert \Omega_{31} \vert ^2 \left(\gamma_2 + 2 \gamma_3\right) \: ,
\\
\gamma_{s} & = & \gamma_1 + \gamma_2 + \gamma_3 \: ,
\\
\Gamma & = & \left(\gamma_1 + \gamma_2\right)^2 \gamma_3 \: .
\end{eqnarray*}
\end{subequations}
Note that in Eq.~(\ref{Floquet_results}), the multiphoton detuning $\Delta$ has been replaced using Eq. (\ref{detuning_equality}).
Also, the contributions without the explicit time dependence via an exponential factor are the direct terms, while the other parts are cross terms.

\subsubsection{\label{non-zero}Non-zero multiphoton detuning ($\Delta \neq 0$)}
In the case of non-zero multiphoton detuning, $\Delta\neq 0 \Leftrightarrow 2\omega_p \neq \omega_c$, only the direct terms in Eqs.~(\ref{Floquet_results}) oscillate in phase with the probe field. Therefore, only these terms contribute to the coherences and thus to the magnetic and electric response. Then, the cross terms in Eqs.~(\ref{macro_response}) vanish and therefore the chirality coefficients vanish, $\tilde{\xi}_{EH} = \tilde{\xi}_{HE} = 0$.
This means that the polarization [magnetization] is entirely determined by the
electric [magnetic] probe field component.
It will turn out in Sec.~\ref{ana_incoherent} that this case is comparable to the incoherently pumped system, in which there are no cross terms either.

\subsubsection{\label{multiphoton_resonance}Multiphoton resonance ($\Delta = 0$)}

We now focus on the case of multiphoton resonance, i.e. $\Delta = 0$ or $\omega_p = \omega_c/2$. Hence, Eqs.~(\ref{eom}) become time-independent and we can now solve the linear system Eq.~(\ref{R_motion}) for a time-independent steady-state solution of $\tilde{R}$ using $\frac{\partial}{\partial t} \tilde{R} = 0$. Now, all terms in Eq.~(\ref{R_ansatz}) contribute, apart from terms that contain $\Omega_{32}$ in higher than first order. From Eqs.~(\ref{second_trafo}) we also find that the simplified relations between the two considered reference frames Eqs.~(\ref{coh_equality}) hold as in the case of incoherent driving.

We again neglect terms of higher order in the probe field Rabi  frequency and arrive at 
\begin{widetext}
\begin{subequations}
\label{resonance_results}
\begin{eqnarray}
\hat{\varrho}_{21} = && - \:  \frac{2 i}{C_{+} D} \left\lbrace8 i \Delta_2^3 \gamma_3 + 4 \Delta_2^2 \left(4 i \Delta_3 - \gamma_s\right) \gamma_3 - \left(4 \Delta_3^2 \gamma_3 + \Gamma\right)\gamma_s 
+ \: 4 \vert \Omega_{31} \vert^2 \left[\gamma_2 \left(\gamma_1 + \gamma_2\right)
\nonumber \right. \right.
\\
&&  \left. \left. + \left(2 i \Delta_3 + \gamma_2\right)\gamma_3 - \gamma_3^2\right] + 2 i \Delta_2 \left[- 4 \vert \Omega_{31} \vert ^2 \left(\gamma_2 - 2\gamma_3\right) + \Gamma + 4 \gamma_3 \Delta_3 \left( \Delta_3  +i  \gamma_s \right) \right] \right\rbrace  \: \Omega_{21} \nonumber
\\
\allowbreak
&&  -   \frac{4\Omega_{31}}{C_{+} D} \left\lbrace 4 \vert \Omega_{31} \vert ^2 \left(\gamma_2 - \gamma_3 \right) - \left[ 2 \left(\Delta_2 +\Delta_3\right) + i \left(\gamma_1 + \gamma_2 \right) \right] \gamma_3 \left( 2 \Delta_2 + i \gamma_s \right) \right\rbrace \: \Omega_{23} \,,
\\
\hat{\varrho}_{32} = && -   \frac{4\Omega_{31}}{C_{-} D}  \left\lbrace4 \vert \Omega_{31} \vert ^2 \left(\gamma_2 - \gamma_3\right) - \left[2 \left(\Delta_2 + \Delta_3\right) + i \left(\gamma_1 + \gamma_2\right)\right]\gamma_3\left(2 \Delta_2 - i \gamma_s\right) \right\rbrace \: \Omega_{12}\nonumber
\\
&&  - \frac{8 i}{C_{-} D} \vert \Omega_{31} \vert ^2 \left[2 i \Delta_3 \gamma_2 +\gamma_3 \left(2i \Delta_2 + \gamma_1 + \gamma_3\right)\right] \: \Omega_{32}\: ,
\end{eqnarray}
\end{subequations}
\end{widetext}
where 

\begin{subequations}
\begin{align*}
C_{\pm} & =  4 \vert \Omega_{31} \vert ^2 +\left(\pm 2 i \Delta_3 + \gamma_3\right) \left( \mp 2 i \Delta_2 + \gamma_s \right) \: ,
\\
D & =  4\left(\Delta_2 + \Delta_3\right)^2 \gamma_3 + \Gamma + 4 \vert \Omega_{31} \vert ^2 \left(\gamma_2 + 2 \gamma_3\right) \:,
\\
\Gamma & =  \left(\gamma_1 + \gamma_2\right)^2 \gamma_3 \: ,
\\
\gamma_s & =  \gamma_1 + \gamma_2 + \gamma_3 \: .
\end{align*}
\end{subequations}
The control field detuning $\Delta_1$ has been eliminated using the relation $\Delta_1 = \Delta_2 + \Delta_3$ which follows from Eq.~(\ref{detuning_equality}) in the case of $\Delta = 0$.

We see that both the direct terms and the cross terms contribute to the coherences and thus to the magnetic and electric response in this case. Therefore, the chirality coefficients are non-zero in the resonance
case $\Delta=0$, see Eqs.~(\ref{macro_response}).

\subsection{\label{ana_incoherent}Incoherently pumped system}

We now consider the electric and magnetic response in the incoherently pumped system shown in  Fig. \ref{fig:closed_loop_scheme}(b). It will serve us as a reference in a comparison to the results of the closed-loop system in order to determine the exact origin of the various contributions to the medium response.

In this system, the transformed Hamiltonian is time-independent. Hence, we can solve for the steady-state solution just as in the case of multiphoton resonance in Sec.~\ref{multiphoton_resonance}. The equations of motion follow from Eqs.~(\ref{eom}) with $\Delta_1 = 0$ and $\Omega_{31} = \Omega_{13} = 0$. Instead of the coherent coupling field, they include an incoherent, bi-directional pump rate $r_1$ on transition $|1\rangle \leftrightarrow |3\rangle$.

Using Eqs.~(\ref{coh_equality}), up to first order in the probe field we find for the coherences in a reference frame oscillating in phase with the probe field:
\begin{subequations}
\label{incoherent_results}
\begin{align}
\hat{\varrho}_{21} = & \frac{2i [r_1 (\gamma_3 - \gamma_2) + (\gamma_1 + \gamma_2) \gamma_3]}{(2 i \Delta_3 + r_1 + \gamma_3)[(\gamma_1 + \gamma_2) \gamma_3 + r_1 (\gamma_2 + 2 \gamma_3)]}\:\Omega_{21} \nonumber
\\
&+\: O(\Omega_{21}^2, \Omega_{32}^2, \Omega_{21}\Omega_{32}) \: ,
\\
\hat{\varrho}_{32} = & \frac{ 2 i r_1 (\gamma_2 - \gamma_3)}{(2 i \Delta_2 + \gamma_s + r_1)[(\gamma_1 + \gamma_2)\gamma_3 + r_1 (\gamma_2 + 2 \gamma_3)]} \: \Omega_{32}\nonumber
\\
&+\: O(\Omega_{21}^2, \Omega_{32}^2, \Omega_{21}\Omega_{32}) \: .
\end{align}
\end{subequations}
In this case, the electric [magnetic] probe transition coherence is determined by the electric [magnetic] probe field component, and no cross-terms appear.

\section{\label{compare}Comparison of the two systems}

We will now proceed with a comparison of the two systems. To this end, we will discuss the expansion coefficients $\hat{\varrho}^{(a,b)}_{32}$ and $\hat{\varrho}^{(a, b)}_{21}$, $(a, b \in \lbrace 0, 1 \rbrace)$ in Eqs.~(\ref{coherence_expansion}), since they determine the magnetic and electric response according to Eqs.~(\ref{final_pol}). 

It will turn out that the direct terms are similar in many regards in both systems. These terms describe the establishment of polarization [magnetization] due to the electric [magnetic] probe field component. But crucial differences are found for the cross terms, which only appear in the closed-loop system for zero multiphoton detuning. These terms characterize the polarization [magnetization] due to the magnetic [electric] probe field component.

\subsection{The direct terms}

\begin{figure}[t]
\centering
\includegraphics[width=4cm]{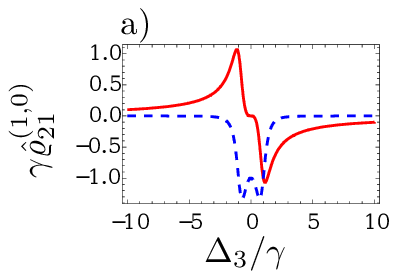}
\hspace*{0.1cm}
\includegraphics[width=4cm]{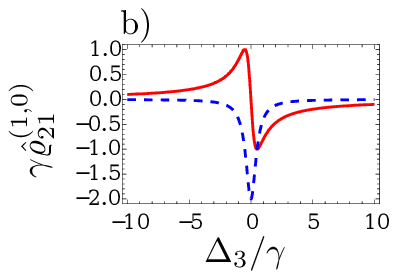}
\\
\includegraphics[width=4cm]{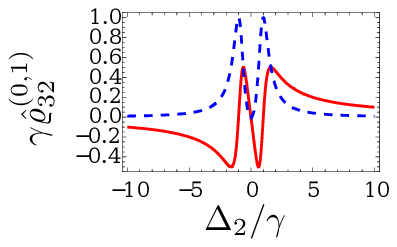}
\hspace*{0.1cm}
\includegraphics[width=4cm]{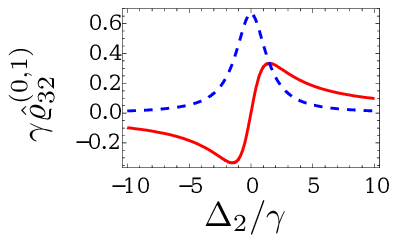}
\caption{\label{fig:rhoCoeffs}(Color online) Real (red solid curve) and imaginary part (blue dashed curve) of the direct terms in the magnetic and electric susceptibility, respectively. The top row shows $\hat{\varrho}_{21}^{(1,0)}$, the bottom row $\hat{\varrho}_{32}^{(0,1)}$. (a) Closed-loop system and (b) incoherently pumped system as shown in Fig.~\ref{fig:closed_loop_scheme}.}
\end{figure}

From Secs.~\ref{non-zero}, \ref{multiphoton_resonance} and \ref{ana_incoherent} it is clear that direct terms appear both in the loop system and in the system with incoherent pumping. In Fig.~\ref{fig:rhoCoeffs}, the corresponding expansion coefficients for the two cases are compared.
Since  transitions $|1\rangle \leftrightarrow |2\rangle$ and $|1\rangle \leftrightarrow |3\rangle$ are electrically dipole-allowed, whereas transition $|2\rangle \leftrightarrow |3\rangle$ is magnetically dipole-allowed, we choose 
the decay rates as $\gamma_1 = \gamma, \gamma_2 = \gamma, \gamma_3 = \alpha^2 \gamma$. For the incoherent case, the pump rate is set to $r_1 = \gamma$, whereas in the closed-loop configuration, the coherent pump field is set to $\Delta_1 = 0, \Omega_{31} = \gamma$. 
All expansion coefficients are plotted against the respective probe field detunings $\Delta_2$ or $\Delta_3$. 

We find that in many respects, the direct terms of both systems behave similarly. Apart from the AC Stark splitting in the closed-loop system, the general structure is comparable and in particular the magnitude of the coefficients is of similar order in both systems. 

\begin{figure}[t]
\centering
\includegraphics[width=0.9\columnwidth]{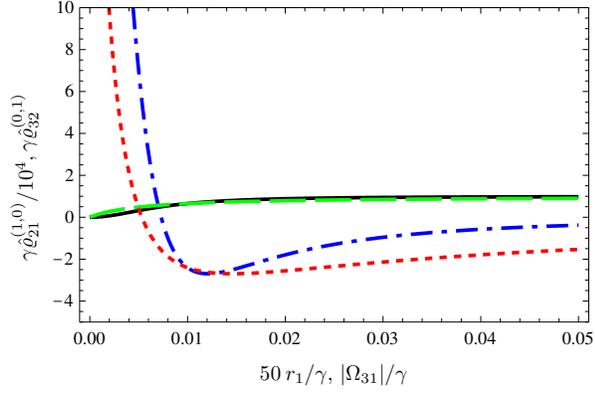}
\caption{\label{fig:pumping}(Color online) Dependence of the expansion coefficients $\hat{\varrho}_{32}^{(0,1)}$ in the closed loop (black solid line) and in the incoherently pumped system (green long dashed line) as well as $\hat{\varrho}_{21}^{(1,0)}$ for the closed-loop (blue dash-dotted) and the incoherently pumped system (red dashed) on the strength of the control field and the incoherent pump rate, respectively. Shown are only the imaginary parts at the maximum of the absolute value of the coefficients. For the incoherent configuration, the maximum of $\hat{\varrho}_{21}^{(1,0)}$ is always at $\Delta_3 = 0$, and the maximum of $\hat{\varrho}_{32}^{(0,1)}$ is at $\Delta_2 = 0$. 
The coefficients are scaled by $\gamma$ to obtain a unitless quantity and  $\hat{\varrho}_{21}^{(1,0)}$ is scaled by a factor of $10^{-4}$.
In this figure, negative values indicate amplification, positive values absorption.}
\end{figure}

The similarities become more apparent when considering the dependence of the direct terms on the coherent pump rate $\Omega_{31}$ and the incoherent pump rate $r_1$, as shown in Fig.~\ref{fig:pumping}.
For this figure, the direct terms of the loop system are evaluated at a particular detuning $\Delta_2$ or $\Delta_3$, respectively, at which the absolute value of the imaginary part becomes maximal. At the same point, 
the real part vanishes. 
For the incoherently pumped system, the corresponding maxima always occur at $\Delta_3=0$ or $\Delta_2=0$, respectively.
This approach allows to compare the two systems independently of the AC Stark splitting appearing in the closed loop system only. Due to the growing splitting with increasing $\vert \Omega_{31} \vert$, a comparison at a fixed detuning would not be meaningful. 
It can be seen from Fig.~\ref{fig:pumping} that both systems show a qualitatively similar dependence on the pumping strength.
The shown imaginary part of the coherences characterizes the absorptive behavior of our systems: positive values stand for absorption and negative values for amplification of the magnetic or electric probe field component. 

Interestingly, in both systems it is possible to choose the respective pump rate in such a way that $\hat{\varrho}_{21}$ vanishes at all frequencies. This is the case at the roots of the dashed lines in Fig.~\ref{fig:pumping}. It turns out that at these points, the populations of states $\vert 1 \rangle$ and $\vert 2 \rangle$ are the same such that the magnetic probe field component can traverse the medium without attenuation and without experiencing diffraction.

For the interpretation of Fig.~\ref{fig:pumping}, we calculate the coherences in terms of the (zeroth order) populations for arbitrary detuning. In the case of incoherent pumping, we obtain
\begin{subequations}
\label{incoherent_coherences}
\begin{eqnarray}
\hat{\varrho}_{21} = && 2 \: \Omega_{21} \frac{ \hat{\varrho}_{11}^{(0)}-\hat{\varrho}_{22}^{(0)}}{2 \Delta_3 - i (r_1 + \gamma_3)} \: ,
\\
\hat{\varrho}_{32} = && 2 \: \Omega_{32} \frac{ \hat{\varrho}_{22}^{(0)}-\hat{\varrho}_{33}^{(0)}}{2 \Delta_2 - i (r_1 + \gamma_1 + \gamma_2 + \gamma_3)} \: .
\end{eqnarray}
\end{subequations}
The zeroth order populations are
\begin{subequations}
\label{incoherent_populations}
\begin{eqnarray}
\hat{\varrho}_{11}^{(0)} = && \frac{(r_1 + \gamma_1 + \gamma_2) \gamma_3}{C} \: ,
\\
\hat{\varrho}_{22}^{(0)} = && \frac{r_1 \gamma_2}{C} \: ,
\\
\hat{\varrho}_{33}^{(0)} = && \frac{r_1 \gamma_3}{C} \: ,
\end{eqnarray}
\end{subequations}
where $C = r_1 \gamma_2 + 2 r_1 \gamma_3 + \gamma_1 \gamma_3 + \gamma_2 \gamma_3$.

Note that in our system, $\gamma_3 \ll \gamma_2$ due to the different multipolarity of the transitions. Therefore, for small pump rates $r_1$, from Eqs.~(\ref{incoherent_populations}) one finds
$\hat{\varrho}_{11}^{(0)} > \hat{\varrho}_{22}^{(0)} > \hat{\varrho}_{33}^{(0)}$. As a result, both probe transitions are absorptive - although for $\hat{\varrho}_{21}$ only in a very small range of $r_1$ as compared to $\gamma$. 

For large $r_1$, Eqs.~(\ref{incoherent_populations}) show that $\hat{\varrho}_{11}^{(0)} < \hat{\varrho}_{22}^{(0)}$ and $\hat{\varrho}_{33}^{(0)} < \hat{\varrho}_{22}^{(0)}$. In fact, $\hat{\varrho}_{11}^{(0)} \approx \hat{\varrho}_{33}^{(0)}$ for $r_1 \gg \gamma$. Hence, the magnetic probe transition becomes amplifying, whereas the electric transition maintains its absorptive character. As expected, from Eqs.~(\ref{incoherent_coherences}), it follows directly that a population inversion causes amplification.

As a side note, we would like to mention the fact that Eqs.~(\ref{incoherent_coherences}) imply vanishing $\hat{\varrho}_{32}$ for $\gamma_2 = \gamma_3$. This is due to the fact that from Eqs.~(\ref{incoherent_populations}), one then finds 
$\hat{\varrho}_{22}^{(0)} = \hat{\varrho}_{33}^{(0)}$.
However, this case is not of relevance for the current analysis, since $\gamma_3 \ll \gamma_2$.

Let us now examine the behavior of the closed-loop system with regard to a change of $\vert \Omega_{31} \vert$. In this case, the coherences are given by
\begin{subequations}
\label{coherent_coherences}
\begin{align}
\hat{\varrho}_{21} &= 2 \: \Omega_{21} \frac{ K_1 (\hat{\varrho}_{22}^{(0)}-\hat{\varrho}_{11}^{(0)}) +4 \vert \Omega_{31} \vert^2 (\hat{\varrho}_{33}^{(0)}-\hat{\varrho}_{11}^{(0)}) }{(4 \vert \Omega_{31} \vert^2 + K_2)K_3} \: ,
\\
\hat{\varrho}_{32} &= 2 \: \Omega_{32} \frac{ K_4 (\hat{\varrho}_{33}^{(0)}-\hat{\varrho}_{22}^{(0)}) + 4 \vert \Omega_{31} \vert^2 (\hat{\varrho}_{33}^{(0)}-\hat{\varrho}_{11}^{(0)}) }{(4 \vert \Omega_{31} \vert^2 + K_5)K_3} \: ,
\end{align}
\end{subequations}
while the populations obey
\begin{subequations}
\label{coherent_populations}
\begin{eqnarray}
\hat{\varrho}_{11}^{(0)} = && \frac{\gamma_3 (K_6 + 4 \vert \Omega_{31} \vert ^2)}{K_7 + 4 \gamma_2 \vert \Omega_{31} \vert ^2 + 8 \gamma_3 \vert \Omega_{31} \vert ^2} \: ,
\\
\hat{\varrho}_{22}^{(0)} = && \frac{4 \gamma_2  \vert \Omega_{31} \vert ^2}{K_7 + 4 \gamma_2 \vert \Omega_{31} \vert ^2 + 8 \gamma_3 \vert \Omega_{31} \vert ^2} \: ,
\\
\hat{\varrho}_{33}^{(0)} = && \frac{4 \gamma_3  \vert \Omega_{31} \vert ^2}{K_7 + 4 \gamma_2 \vert \Omega_{31} \vert ^2 + 8 \gamma_3 \vert \Omega_{31} \vert ^2}
\end{eqnarray}
\end{subequations}
where the $K_l, (l \: \in \: \lbrace1, \dots 7 \rbrace)$ are coefficients independent of $\Omega_{31}$.

Again, the coherence $\hat{\varrho}_{ij}$ is determined by the difference of the populations of $\hat{\varrho}_{ii}^{(0)}$ and $\hat{\varrho}_{jj}^{(0)}, \: [(i,j) \: \in \: \lbrace (2,1),(3,2)]$. For small $\vert \Omega_{31} \vert$, the term proportional to $\hat{\varrho}_{33}^{(0)}-\hat{\varrho}_{11}^{(0)}$ can be neglected. For large $\vert \Omega_{31} \vert$, this term outweighs the others at first sight. However, for $\vert \Omega_{31} \vert \rightarrow \infty$ and arbitrary, but fixed detunings, one finds
\begin{subequations}
\label{coherent_pop_limits}
\begin{eqnarray}
\hat{\varrho}_{11}^{(0)} \rightarrow && \frac{1}{\frac{\gamma_2}{\gamma_3}+2} \: ,
\\
\hat{\varrho}_{22}^{(0)} \rightarrow && \frac{1}{2\frac{\gamma_3}{\gamma_2}+1} \: ,
\\
\hat{\varrho}_{33}^{(0)} \rightarrow && \frac{1}{\frac{\gamma_2}{\gamma_3}+2} \: .
\end{eqnarray}
\end{subequations}
Therefore, $\hat{\varrho}_{11}^{(0)}-\hat{\varrho}_{33}^{(0)} \rightarrow 0$ such that also in this case, $\hat{\varrho}_{ij}$ is determined by $\hat{\varrho}_{ii}^{(0)} - \hat{\varrho}_{jj}^{(0)}$.

Eqs. (\ref{coherent_pop_limits}) also imply that for a strong control field, both probe transitions show opposite absorptive behavior (see Fig. \ref{fig:pumping}): absorption in case of the upper probe transition, and amplification for the lower one. The reason is that due to the small decay rate $\gamma_3$, most population is trapped in state $|2\rangle$, such that the relevant population differences in Eqs.~(\ref{coherent_coherences}) have opposite sign.
For small $\vert \Omega_{31} \vert$, Eqs.~(\ref{coherent_populations}) yield $\hat{\varrho}_{11}^{(0)} > \hat{\varrho}_{22}^{(0)} > \hat{\varrho}_{33}^{(0)}$, which explains the absorptive properties in Fig.~\ref{fig:pumping}. However, the first inequality is not obvious and can only be deduced by an exact knowledge of $K_6$.

\begin{figure}
\centering
\includegraphics[width=7cm]{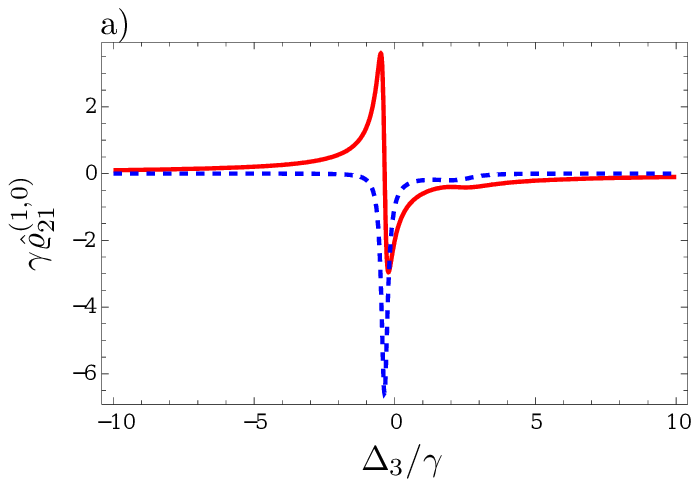}
\includegraphics[width=7cm]{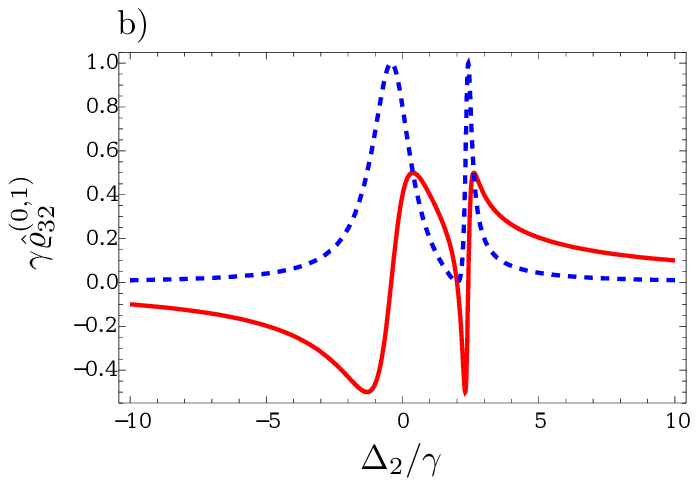}
\caption{\label{fig:rhoCoeffs_Delta}(Color online) Real (red solid line) and imaginary part (blue dashed curve) of the expansion coefficients (a) $\hat{\varrho}_{21}^{(1,0)}$ and (b) $\hat{\varrho}_{32}^{(0,1)}$ in Eqs.~(\ref{coherence_expansion}). The curves are drawn  for $\Delta_1 = 2 \gamma$ in the closed-loop system. The coefficients are scaled by $\gamma$. The parameters are as in Fig.~\ref{fig:rhoCoeffs}(a), except for the non-vanishing detuning of the control field.}
\end{figure}

Before we come to the discussion of the cross terms, we would like to note that, of course, there are also differences between the direct terms of the two systems. We already discussed the AC Stark shift that occurs only in the loop system.
But the closed-loop system also offers more degrees of freedom than the incoherently pumped system, most importantly $\Delta_1$. The role of the control field detuning $\Delta_1$ is shown in Fig.~\ref{fig:rhoCoeffs_Delta}. In this figure, we plot the expansion coefficients of the direct terms in the coherence over the probe field detunings $\Delta_2$ and $\Delta_3$ for $\Delta_1 = 2 \gamma$.
The detuning $\Delta_1$ essentially determines the position of one of the
maxima of the imaginary part of the response function. Increasing $\Delta_1$ moves the peak to higher frequencies, while decreasing it moves it to lower frequencies. For example, in Fig.~\ref{fig:rhoCoeffs_Delta}, the corresponding maxima can be seen close to $\Delta_2=2\gamma$.

\subsection{The cross terms}

\begin{figure*}[t]
\centering
\includegraphics[height=4.5cm]{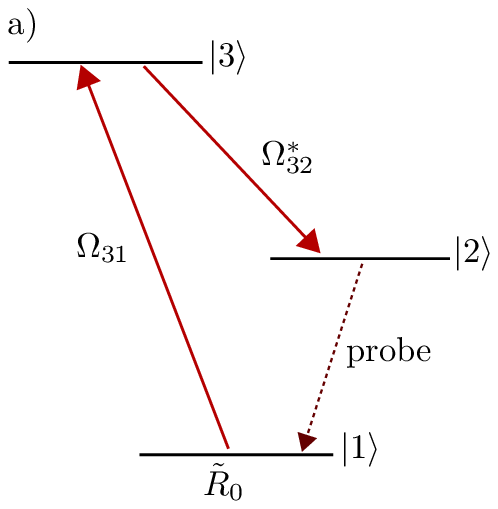}
\hspace*{0.5cm}
\includegraphics[height=4.5cm]{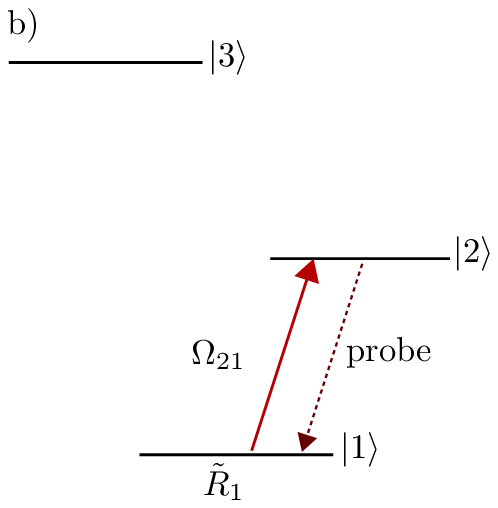}
\hspace*{0.5cm}
\includegraphics[height=4.5cm]{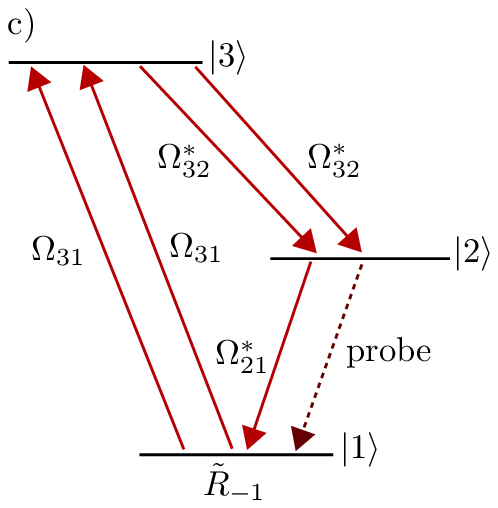}
\caption{\label{fig:R_mag}(Color online) Interpretation of the different scattering processes that contribute to $\tilde{\varrho}_{21}$ in the case of $\Delta = 0$. (a) $\tilde{R}_{0}$ contributes to the cross term in $\tilde{\varrho}_{21}$, and thus to the chirality $\tilde{\xi}_{HE}$. (b) $\tilde{R}_{1}$ contributes to the direct term, i.e. to the magnetic susceptibility. (c) $\tilde{R}_{-1}$ is of higher order in either one of the probe field Rabi frequencies and therefore negligible in our calculation.}
\end{figure*}

\begin{figure*}[t]
\centering
\includegraphics[height=4.5cm]{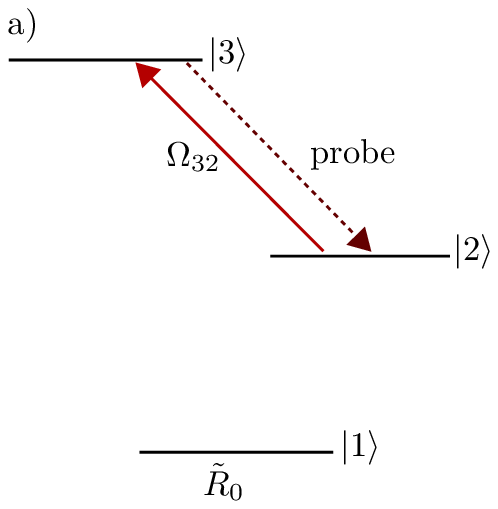}
\hspace*{0.5cm}
\includegraphics[height=4.5cm]{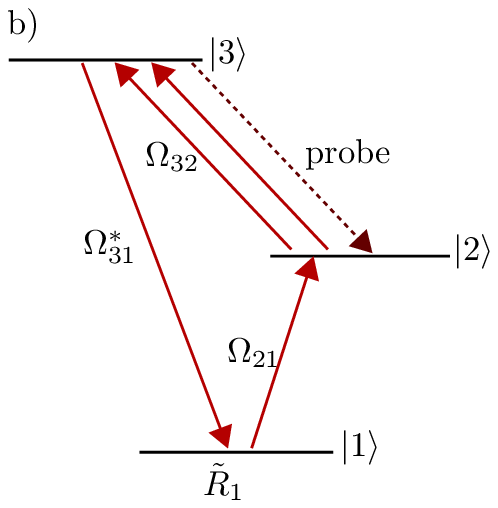}
\hspace*{0.5cm}
\includegraphics[height=4.5cm]{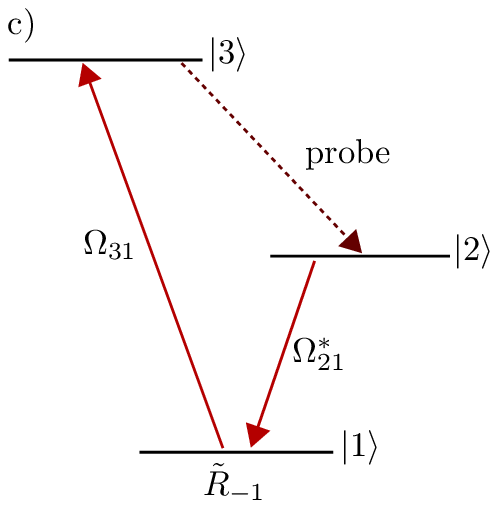}
\caption{\label{fig:R_el}(Color online) Interpretation of the different  scattering processes that contribute to $\tilde{\varrho}_{32}$ in the case of $\Delta = 0$. (a) $\tilde{R}_{0}$ contributes to the direct term in $\tilde{\varrho}_{32}$, and thus to the electric susceptibility. (b) $\tilde{R}_{1}$ is of higher order in either one of the probe field Rabi frequencies and can therefore be neglected. (c) $\tilde{R}_{-1}$ contributes to one of the cross terms, i.e., to one of the chirality coefficients.}
\end{figure*}

\begin{figure}[t]
\centering
\includegraphics[width=7cm]{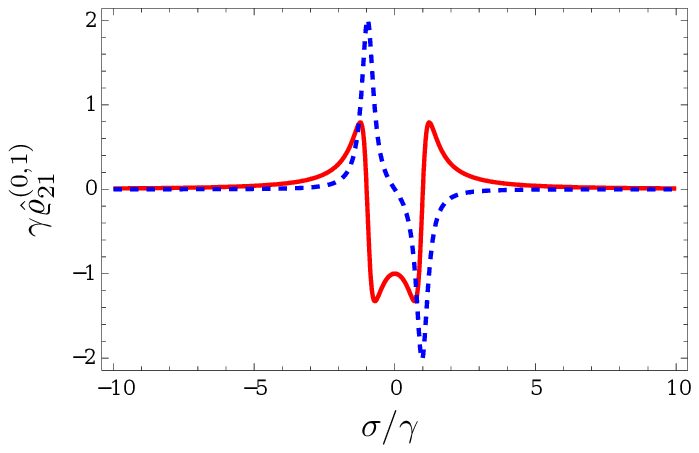}
\caption{\label{fig:chiralities}(Color online) Real (red solid line) and imaginary part (blue dashed) of the expansion coefficient $\hat{\varrho}_{21}^{(0,1)}$ in the closed-loop system [see  Eqs.~(\ref{coherence_expansion})]. $\hat{\varrho}_{32}^{(1,0)}$ is not shown, since it is virtually identical to the shown results. The coefficients correspond to cross terms and determine $\tilde{\xi}_{HE}$ and $\tilde{\xi}_{EH}$, respectively. The plotted coefficients are phase-dependent; in this figure all phases are set to zero.
The variable $\sigma = (\Delta_3 - \Delta_2)/2 = \omega_2 - (\omega_3 + \omega_1)/2$ denotes the shift of the eigenfrequency of $\vert 2 \rangle$ with respect to the average frequency of $\vert 1 \rangle$ and $\vert 3 \rangle$. Then, for all values of $\sigma$, the multiphoton resonance condition $\Delta=0$ is fulfilled for a fixed probe field frequency $\omega_p$.}
\end{figure}

Let us now come to the most important difference between the two systems: The coherences of the closed-loop system have cross terms, while the coherences of the incoherently pumped system do not. However, as found in Sec.~\ref{closed_loop_system}, 
the cross terms only contribute to the magnetic and electric response for $\Delta = 0$, i.e. for $\omega_p = \omega_c/2$. The following comparison serves as basis for the conclusion which will be drawn in the discussion section regarding the enhancement of the magnetic response.

The mathematical origin of the cross terms can be identified from the derivation of the coherences in Section~\ref{closed_loop_system}. The relevant response of the system to the probe field is given by the contributions of the respective probe transition coherences oscillating in phase with the incident probe field.
According to this criterion, for $\Delta \neq 0$, only one of the terms in Eq.~(\ref{R_ansatz}) contributes to each probe field coherence.
In contrast, for $\Delta = 0$, all terms in Eq.~(\ref{R_ansatz}) contribute to this response. The additional terms lead to the cross terms discussed here. In principle, also other terms contribute in this case, but they are of higher order in the probe field Rabi frequencies and can therefore be neglected in linear response theory.

How can we interpret the different contributions $\tilde{R}_{k}$ $(k \: \in \: \lbrace-1,0,1\rbrace)$ to Eq.~(\ref{R_ansatz}) physically? An explicit calculation reveals their dependence on the probe and control field Rabi frequencies. The control field Rabi frequencies appear for two different reasons. First, the populations depend on the control field Rabi frequencies. But second, the Rabi frequencies also indicate the physical process described by the respective terms.
The obtained combinations of the different Rabi frequencies lead to an interpretation of $\tilde{R}_{k}$ as depicted in Figs.~\ref{fig:R_mag} and \ref{fig:R_el}. 
For example, $\tilde{R}_{0}$ contributing to  $\tilde{\varrho}_{21}$ is shown 
in Fig.~\ref{fig:R_mag}(a). This term arises from the scattering of the control field off of transition $|1\rangle \to |3\rangle$ and the probe field off of transition  $|3\rangle \to |2\rangle$ into the probe transition $|2\rangle \to |1\rangle$, which contributes to the magnetic response.

We now turn to a numerical study of the cross terms.
In Fig.~\ref{fig:chiralities},  we plot $\hat{\varrho}_{21}^{(0,1)}$, multiplied by a factor of $\gamma$ to achieve unitless quantities. $\hat{\varrho}_{32}^{(1,0)}$ is not shown, as it is virtually identical for the chosen parameters. 
It is important to note that in Fig.~\ref{fig:chiralities}, the cross terms are plotted over the variable $\sigma$ which is defined as $\sigma = \frac{1}{2} (\Delta_3 - \Delta_2) = \omega_2 - \frac{1}{2}(\omega_3 + \omega_1)$. This new variable can be interpreted as the energy shift of state $\vert 2 \rangle$ with respect to the average energy of $\vert 1 \rangle$ and $\vert 3 \rangle$. Hence, in a plot against $\sigma$, effectively state $\vert 2 \rangle$ is moved. In this way, the probe field frequency remains fixed such that the multiphoton resonance condition $\Delta = 0 \Leftrightarrow \omega_p = \omega_c/2$ is fulfilled for all values of $\sigma$. For $\sigma = 0$, $\vert 2 \rangle$ lies in the very middle of $\vert 1 \rangle$ and $\vert 3 \rangle$.

It turns out that apart from the phases of the dipole moments, the chiralities in Eqs.~(\ref{macro_response}) depend on the phase 
$\psi-2\phi+ \vec{K}\vec{r}$ arising from the closed interaction loop, where $\vec{K}=(  \vec{k}_c -2\vec{k}_p )$ is the so-called wave vector mismatch. 
For the plot in Fig.~\ref{fig:chiralities}, we set all involved phases to zero. This does not affect our final conclusion regarding the enhancement of the magnetic response, since it will only be based on the magnitude of the cross terms. If in an experiment the difference between the absolute field phases $\psi-2\phi$ is not fixed, then the chiralities average to zero. While absolute phase control is very difficult to achieve, relative phase control has been accomplished experimentally~\cite{PhysRevA.59.2302}. In related systems, the phase dependence of the cross terms can be made independent of the probe field phase, as will be discussed in Sec.~\ref{discussion}.
The observed phase dependence is a characteristic of closed loop systems, and has been observed in related systems as well~\cite{evers, orth:paper}. The phase can be calculated by following the interaction loop, and counting phases of fields that deexcite the atom and phases of fields that excite the atom throughout this loop path with opposite sign.

Due to the dependence of the chiralities on $\vec{K}\vec{r}$, a further condition for the enhancement arises, namely, that the so-called wave vector mismatch should vanish, i.e. $\vec{K} = 0$. This condition on the relative propagation directions of the different fields for the enhancement to take place can be fulfilled, for example, for co-propagating fields~\cite{evers}.

\subsection{Enhancement of the magnetic response}
We are now in the position to evaluate the magnitude of the magnetic response. For this, we examine the different contributions in Eq.~(\ref{final_mag}).
In particular, we compare the two contributions to the magnetization, which except for the common prefactor are given by
\begin{subequations}
\begin{align}
M_1 &= d_{32}  \mu_{21} \: \mathcal{E}_p \: e^{i\Phi} \: \hat{\varrho}_{21}^{(0,1)}\,,
\\
\label{M2}
M_2 &= \mu_{21}^{2} \: \mathcal{B}_p \: \hat{\varrho}_{21}^{(1,0)}\,.
\end{align}
\end{subequations}
Here, $M_1$ refers to the cross term contribution that only contributes at $\Delta = 0$ and $M_2$ denotes the direct term.

First, Figs.~\ref{fig:rhoCoeffs} and \ref{fig:chiralities} show that in the closed loop system, 
\begin{equation}
\label{expcoeff}
\vert \hat{\varrho}_{21}^{(1,0)} \vert \approx \vert \hat{\varrho}_{21}^{(0,1)} \vert\,.
\end{equation}
Thus, the magnitude of the two expansion coefficients is comparable. 
The order of magnitude of the involved transition dipole moments can be estimated as,
\begin{subequations}
\label{estimation1}
\begin{eqnarray}
\mu_{21} & \sim & \mu_B \sim e a_0 \alpha c \: ,
\\
d_{32} & \sim & e a_0 \: ,
\end{eqnarray}
\end{subequations}
where $\mu_B$ is the Bohr magneton, $e$ the elementary charge, $a_0$ the Bohr radius, $\alpha$ the fine-structure constant and c the vacuum speed of light. Finally, 
\begin{equation}
\label{estimation2}
\mathcal{B}_p = \frac{1}{c} \mathcal{E}_p \: .
\end{equation}
With Eqs. (\ref{estimation1}) and (\ref{estimation2}), we thus arrive at
\begin{equation}
\vert M_1 \vert \approx \alpha^{-1} \vert M_2 \vert\,.
\end{equation}
This means that at multiphoton resonance, the magnetic response of the closed-loop system is enhanced by a factor of $\alpha^{-1}$ due to the scattering of the electric probe field component into the magnetic probe transition (described by $\tilde{R}_0$ in Fig. \ref{fig:R_mag}).
A similar argument shows that the direct terms of the incoherently pumped and the closed loop system are comparable in magnitude, such that the closed-loop system in multiphoton resonance also allows an enhancement of the magnetic response by $\alpha^{-1}$ as compared to the incoherently pumped system.

\begin{figure}[bt]
\centering
\includegraphics[width=4cm]{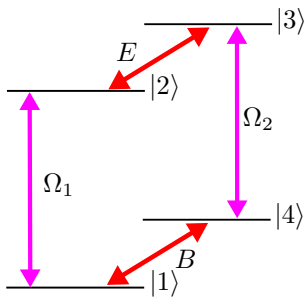}
\caption{\label{fig:4level}(Color online) Example level scheme for
magnetic response enhancement independent of the probe field frequency.
$\Omega_1$ and $\Omega_2$ are coherent coupling fields with frequencies $\omega_1$ and $\omega_2$, respectively. $E$ and $B$ are the electric and magnetic component of the probe field with frequency $\omega_p$. Here, the
closed loop path contains an absorption and an emission of a probe field
photon such that the  multiphoton detuning $\Delta = \omega_1 - \omega_2$ can
be satisfied for arbitrary probe field frequencies.}
\end{figure}

\section{\label{discussion}Discussion}

We have seen that the direct response, which describes the polarization [magnetization] created by the electric [magnetic] probe field component, is of the same order of magnitude in all three cases considered in Sec.~\ref{analytical_results}.
Only at multiphoton resonance, the response of the closed-loop system in addition contains a term corresponding to the scattering of the electric probe field component into the magnetic probe field transition. We could show that this scattering effectively enhances the magnetic response by one inverse power of the fine structure constant $\alpha^{-1}$, and thus clearly identify the mechanism leading to the enhanced magnetic response.

Of course, the reverse process of scattering the magnetic field into the electric probe field mode can also occur. However, Eqs.~(\ref{final_pol}) show that this does not lead to an enhancement of the electric response, since, mathematically speaking, $\vert d_{32}^2 \: \vec{E}_p \vert \sim \alpha^{-1} \vert d_{32} \: \mu_{32} \: \vec{B}_p \vert$. Physically speaking, the coupling of the magnetic probe field component to a magnetic transition is smaller than the coupling of the electric probe field component to an electric transition by a factor of $\alpha$.

The multiphoton resonance condition restricts the magnetic enhancement in the system discussed here to a single probe field frequency. Extended closed-loop systems can be constructed in such a way that a complete loop contains an excitation on the electric probe transition and a deexcitation on the magnetic probe transition, or vice versa~\cite{fleischhauer,orth:paper}.  A simple example for this is shown in Fig.~\ref{fig:4level}. In such a system, the multiphoton resonance condition does not depend on the frequency of the probe field, such that it can be fulfilled for arbitrary probe field frequencies by suitably choosing the coupling field frequencies. The physical interpretation identified in our analysis directly carries over to these extended systems.

While the parametric enhancement by $\alpha^{-1}$ is universal, other closed-loop systems could in principle lead to a modification of the relative magnitude of the different expansion coefficients in Eq.~(\ref{expcoeff}).
If this ratio can be altered favorably, then the enhancement can be even higher. It remains to be seen, however, whether there is a physical mechanism that enables one to change this ratio to a great extend.

In conclusion, we have studied a mechanism for the enhancement of the magnetic response in atomic systems. We found that the studied enhancement can be traced back to a scattering of the electric probe field component into the magnetic probe field component. This scattering is only possible in closed-loop systems, where the applied laser fields form a closed interaction loop. Our analysis shows that the magnetic field enhancement is of order $\alpha^{-1}$ and occurs if the multiphoton resonance condition is fulfilled.


\begin{thebibliography}{10}

\bibitem{veselago}
V.~G. Veselago, Sov. Phys. Usp \textbf{10}, 509 (1968).

\bibitem{review1}
V.~Veselago, L.~Braginsky, V.~Shkliver, and C.~Hafner, J. Comput. Theor.
  Nanosci. \textbf{3}, 189 (2006).

\bibitem{Sh2007}
V.~M. Shalaev, Nature Photonics \textbf{1}, 41 (2007).

\bibitem{D.R.Smith08062004}
D.~R. Smith, J.~B. Pendry, and M.~C.~K. Wiltshire, Science \textbf{305}, 788
  (2004).

\bibitem{pendry:lens}
J.~B. Pendry, Phys. Rev. Lett. \textbf{85}, 3966 (2000).

\bibitem{R.A.Shelby04062001}
R.~A. Shelby, D.~R. Smith, and S.~Schultz, Science \textbf{292}, 77 (2001).

\bibitem{soukoulis}
E.~Cubukcu, K.~Aydin, E.~Ozbay, S.~Foteinopoulou, and C.~M. Soukoulis, Nature
  \textbf{423}, 604 (2003).

\bibitem{PhysRevLett.90.137401}
A.~A. Houck, J.~B. Brock, and I.~L. Chuang, Phys. Rev. Lett. \textbf{90},
  137401 (2003).

\bibitem{eleftheriades}
A.~Grbic and G.~V. Eleftheriades, Phys. Rev. Lett. \textbf{92}, 117403 (2004).

\bibitem{zhang:137404}
S.~Zhang, W.~Fan, N.~C. Panoiu, K.~J. Malloy, R.~M. Osgood, and S.~R.~J.
  Brueck, Phys. Rev. Lett. \textbf{95}, 137404 (2005).

\bibitem{CostasM.Soukoulis01052007}
C.~M. Soukoulis, S.~Linden, and M.~Wegener, Science \textbf{315}, 47 (2007).

\bibitem{oktel}
M.~O. \"Oktel and O.~E. M\"ustecaplioglu, Phys. Rev. A \textbf{70}, 053806
  (2004).


\bibitem{mandel}
Q.~Thommen and P.~Mandel, Phys. Rev. Lett. \textbf{96}, 053601 (2006).

\bibitem{fleischhauer}
J.~K\"astel, M.~Fleischhauer, S.~F. Yelin, and R.~L. Walsworth, Phys. Rev.
  Lett. \textbf{99}, 073602 (2007).

\bibitem{orth:paper}
P.~P. Orth, J.~Evers, and C.~H. Keitel, arXiv:0711.0303 (2007).


\bibitem{PhysRevA.46.1468}
M.~Fleischhauer, C.~H. Keitel, M.~O. Scully, C.~Su, B.~T. Ulrich, and S.-Y.
  Zhu, Phys. Rev. A \textbf{46}, 1468 (1992).

\bibitem{pendry:chirality}
J.~B. Pendry, Science \textbf{306}, 1353 (2004).


\bibitem{Buckle}
S.~J. Buckle, S.~M. Barnett, P.~L. Knight, M.~A. Lauder, and D.~T. Pegg, Opt.
  Acta \textbf{33}, 2473 (1986).

\bibitem{Kosachiov}
D.~V. Kosachiov, B.~G. Matisov and Y.~V. Rozhdestvensky, J. Phys. B \textbf{25}, 2473
  (1992).

\bibitem{PhysRevLett.70.3243}
W.~E. van~der Veer, R.~J.~J. van Diest, A.~D\"onszelmann, and H.~B. van Linden
  van~den Heuvell, Phys. Rev. Lett. \textbf{70}, 3243 (1993).

\bibitem{PhysRevA.53.3444}
W.~Maichen, F.~Renzoni, I.~Mazets, E.~Korsunsky, and L.~Windholz, Phys. Rev. A
  \textbf{53}, 3444 (1996).

\bibitem{PhysRevA.59.2302}
E.~A. Korsunsky, N.~Leinfellner, A.~Huss, S.~Baluschev, and L.~Windholz, Phys.
  Rev. A \textbf{59}, 2302 (1999).

\bibitem{PhysRevA.60.4996}
E.~A. Korsunsky and D.~V. Kosachiov, Phys. Rev. A \textbf{60}, 4996 (1999).


\bibitem{PhysRevLett.84.5308}
A.~J. Merriam, S.~J. Sharpe, M.~Shverdin, D.~Manuszak, G.~Y. Yin, and S.~E.
  Harris, Phys. Rev. Lett. \textbf{84}, 5308 (2000).

\bibitem{hinze}
S.~A. Babin, S.~I. Kablukov, U.~Hinze, E.~Tiemann, and B.~Wellegehausen, Opt.
  Lett. \textbf{26}, 81 (2001).


\bibitem{PhysRevA.66.053409}
G.~Morigi, S.~Franke-Arnold, and G.-L. Oppo, Phys. Rev. A \textbf{66}, 053409
  (2002).


\bibitem{PhysRevLett.93.223601}
A.~F. Huss, R.~Lammegger, C.~Neureiter, E.~A. Korsunsky, and L.~Windholz, Phys.
  Rev. Lett. \textbf{93}, 223601 (2004).

\bibitem{PhysRevLett.93.190502}
V.~S. Malinovsky and I.~R. Sola, Phys. Rev. Lett. \textbf{93}, 190502 (2004).

\bibitem{shpaisman:043812}
H.~Shpaisman, A.~D. Wilson-Gordon, and H.~Friedmann, Physical Review A
  \textbf{71}, 043812 (2005).

\bibitem{evers}
M.~Mahmoudi and J.~Evers, Phys. Rev. A \textbf{74}, 063827 (2006).

\bibitem{kajari-schroder:013816}
S.~Kajari-Schroder, G.~Morigi, S.~Franke-Arnold, and G.-L. Oppo, Physical
  Review A \textbf{75}, 013816 (2007).

\bibitem{jackson}
J.~D. Jackson, \emph{Classical Electrodynamics} (Wiley, New York, 1998).

\bibitem{ScZu1997}
M.~O. Scully and M.~S. Zubairy, \emph{Quantum Optics} (Cambridge University
  Press, Cambridge, 1997).

\bibitem{floquet}
M.~G. Floquet, Ann. Sci. Ec. Normale Super. \textbf{12}, 47 (1883).


\end{thebibliography}

\end{document}